\documentclass [12pt]{article}
\def\maj#1{\ifmmode\mbox{\usefont{U}{msb}{m}{n}#1}\else{\usefont{U}{msb}{m}{n}#1}\fi}

\makeatletter \@addtoreset{equation}{section} \makeatother

\usepackage{graphics,epsfig,graphicx}
\usepackage{color}

\begin{document}

\title{\textbf{Dressed atom versus exciton polariton:\\ From Rabi oscillations to \\ the Fermi Golden rule}}
\author{Francois  Dubin, Monique Combescot, Bernard Roulet
 \\ \textit{GPS, Universit\'e Pierre et Marie
Curie, CNRS}\\
\textit{Campus Boucicaut, 140 rue de Lourmel, 75015 Paris}}
\date{}
\maketitle


\maketitle

\begin{abstract}
We rederive the dressed atom and the exciton polariton within the
{\it same} framework to make clear that their difference only
comes from the number of electrons available for photoexcitations.
Using it, we analytically show how the time dependence of the
photon number transforms from an oscillating behavior (at the
stimulated or vacuum Rabi frequency) to an exponential decay
(identical for atom and semiconductor) when the excited state
lifetime decreases. Although the matter ground state is in both
cases coupled by monochromatic photons {\it not to a continuum but
to a discrete state}, this decay yet follows a kind of Fermi
Golden rule. The energy conservation it contains, is however
conceptually different.
\end{abstract}

\newpage

A very large amount of our physical understanding comes from
matter-photon interaction, this matter being either a dilute set
of atoms as in atomic physics, or a dense arrangement as in solid
state physics. In all cases, the photons induce a coupling between
electronic levels. It is known that for a single two-level atom,
the electron Rabi oscillates \cite{ref1} between its two possible
levels while for solids, the photons are absorbed with a
transition rate given by the Fermi Golden rule \cite{ref2}. The
reason invoked for this behavior change is the energy distribution
of the states coupled by photons to the matter ground state: In
usual solids, their energies are close to a continuum, so that
Rabi oscillations are destroyed by interferences. These
interferences in fact lead to an overall exponential decay
controlled by excited states which have an energy close to the
initial one. The energy conservation appears through a delta
"function", $\delta_{t}(\epsilon)=sin(\epsilon t/2)/\pi\epsilon$,
which has a width, $t^{-1}$ (if $\hbar$=1), equal to  \cite{ref3}
the characteristic energy of the Heisenberg uncertainty principle.

A case of special interest is however the interaction of photons
with semiconductors, because the hole left in the valence band can
form bound states with the photocreated electron, called excitons.
In this case, the states coupled by photons to the semiconductor
ground state do not form a continuum, as in usual solids, because
the photocreated exciton has a well defined center of mass
momentum, its components being the ones of photon along the
exciton free directions. Consequently, the standard reason for
having a photon absorption given by the Fermi Golden rule, can not
be invoked in the case of exciton formation.

In addition, long ago, Hopfield has shown that photons  and
excitons form mixed states, called polaritons \cite{ref4}. Being
eigenstates of the coupled semiconductor-photon hamiltonian, they
cannot decay, so that no photon absorption can result from the
semiconductor-photon interaction alone, whatever the strength of
this interaction is. According to Hopfield, the experimentally
observed absorption is due to additional couplings of the exciton
component of the polariton to external reservoirs \cite{b}.

Cases in which the polariton picture has to be used and cases in
which photons are barely absorbed according to the Fermi Golden
rule, are said to correspond to "strong" and "weak" couplings
\cite{ref5,ref6}. Although the change from one regime to the other
has been correctly related \cite{ref7} to the strength of the
semiconductor-photon interaction compared to the exciton
broadening induced by external couplings, to our best knowledge,
no direct derivation of a photon {\it absorption} for transitions
to a {\it discrete} state has been given. More precisely, how,
starting from the Hopfield's polariton, is constructed the
exponential decay corresponding to a photon absorption when the
ground state is coupled to {\it one} state only? In particular, is
the characteristic time of this decay really given by the Fermi
Golden rule, as implicitly assumed by everyone, since the initial
state is not coupled to a continuum, so that the energy
conservation appearing in this rule cannot be due to destructive
interferences between final states close in energy?

We wish to stress that the problems linked to strong versus weak
couplings are usually discussed in the context of photon {\it
emission} from excitons in a quantum well, either in free space or
in a microcavity \cite{ref7bis,7bb}. The physics is then totally
different. The possible momenta of photons {\it emitted} by an
exciton with a momentum $\textbf{Q}_{//}$ along the well, are
$\textbf{Q}_{//}$+$\textbf{q}_z$ with
E$_x$+$\hbar^2\textbf{Q}^2_{//}$/2$M_x$ =
$\hbar$v$|\textbf{Q}_{//}+\textbf{q}_z|$. For a quantum well in
vacuum, there is a continuous set of $\textbf{q}_z$ which fulfill
this relation. The exciton being then coupled to a {\it continuum}
of photons, the photon emission transition rate has to be given by
the Fermi Golden rule. Because, in the Fermi Golden rule, the
exciton-photon interaction enters at lowest order only, these
excitons seem to have a "weak coupling" with photons. On the
opposite, when the well is in a microcavity, the possible cavity
modes $q_z$ are discrete. The exciton being no more coupled to a
continuum, there is no reason for the photon emission to follow
the Fermi Golden rule. And indeed, in order to explain the
observed results, the exciton-photon coupling has be treated
exactly, through the polariton, so that these excitons seem to
have a "strong coupling" with photons \cite{notea}. Such a photon
{\it emission} has to be contrasted with the photon {\it
absorption} considered here, in which one photon
$\textbf{q}=\textbf{q}_{//}+\textbf{q}_z$ is coupled to {\it one}
exciton only, its momentum being $\textbf{q}_{//}$: As the final
state is then discrete, the reason for a possible regime in which
the Fermi Golden rule is valid, is a priori not obvious.

In this letter, we address this quite fundamental question of
semiconductor physics: why can we use the Fermi Golden rule for
photon {\it absorption} in a semiconductor since the final state
is discrete.

In order to tackle this question, we follow a dressed atom
approach \cite{ref8,ref9}: we consider N photons interacting
either with a single atom or a semiconductor, initially in their
ground state. We look for the time evolution of the photon number,
$\mathcal{N}$(t), in terms of the matter-photon coupling, the
excited state detuning and its broadening. The problem is actually
more tricky for semiconductors than for atoms, because, the
valence band having very many electrons, all photons can a priori
be transformed into excitations so that the dimension of the
coupled subspace is not (1+1), as for a dressed atom, but (1+N).
By using {\it generalized polariton operators dressed by the
exciton external couplings}, it is however possible to obtain an
analytical expression of the photon number time evolution, also in
the semiconductor case.

We recover that, when the excited state broadening $\gamma$ is
small, $\mathcal{N}$(t) oscillates with the vacuum Rabi frequency,
$\Omega_1$, in the case of semiconductor and the stimulated one,
$\Omega_N$=$\Omega_1\sqrt{N}$, in the case of atom. On the
opposite, when $\gamma$ is large, $\mathcal{N}$(t) for
semiconductor and atom exhibits the {\it same} exponential decay,
with a characteristic time given by a kind of Fermi Golden rule.
This decay, which cannot be due to standard destructive
interferences between Rabi oscillations of quasi-continuous
states, has a quite different origin: The energy conservation it
contains, is not the usual $\delta_t(\epsilon)$, but
$\widehat{\delta}_\gamma(\epsilon)=\gamma/\pi(\epsilon^2+\gamma^2)$.
This other delta "function" has a $t$ independent width, $\gamma$.
For large $t$, more precisely for $\gamma t\gg1$, its width is in
fact larger than the width, $t^{-1}$, of the standard rule, as
fully reasonable since the energy uncertainty due to the excited
state broadening $\gamma$ is then larger the one of the Heisenberg
principle, the energy conservation being possible to enforce at
the larger of the two.

\section{Two-level atom dressed by photons}

The presentation we give here of this quite well known problem
\cite{ref8,ref9}, allows to enlighten the similarities and
differences between atom and exciton.

In a two-level atom, the photon transfers the electron from level
0 to level 1, leaving a hole in the level 0 (see fig.(1a)). Once
in level 1, the electron can return to level 0 either by emitting
the same photon or by relaxation such as the emission of
fluorescent photons. The coupled atom-photon hamiltonian can be
written as
\begin{equation}\label{eq1}
H= \omega_{p}A^{+}A +
(\omega_{x}-i\gamma)B^{+}B+(\mu^{*}B^{+}A+h.c.)
\end{equation}
where $\omega_p$ is the photon energy, $\omega_x$ the energy
difference between levels 1 and 0 and $\gamma=1/2\tau$ the
broadening of level 1 induced by its relaxation. $\mu^*$ is the
matrix element for the transformation of one photon into one
excitation. $A^+$ creates one photon while $B^+$ creates one
excitation. In the case of a single two-level atom, it is such
that $B^{+} \mid X^{0}\rangle =\mid X^1 \rangle$, $B \mid X^{1}
\rangle= \mid X^{0}\rangle$, while $B^{+} \mid X^{1}\rangle$=$B
\mid X^{0}\rangle =0$ where $\mid X^1 \rangle$ and $\mid X^0
\rangle$ correspond to the electron in the levels 1 and 0,
respectively.

Let us consider $\mid \psi_{N}(t=0)\rangle=\mid N,X^{0} \rangle$
as initial state: the system has N photons, the atom being in its
ground state. The atom-photon interaction couples this state to
$\mid N-1,X^{1} \rangle$. The H eigenvalues in this restricted
subspace are
\begin{equation}\label{eq2}
E^{(\pm)}_N= N\omega_{p}+ (\widetilde{\Delta} \pm
\widetilde{\Omega}_{N})/2
\end{equation}
where $\widetilde{\Delta}=\Delta-i\gamma$ with
$\Delta=\omega_{x}-\omega_{p}$ being the detuning while
$\widetilde{\Omega}_{N}^{2}=\widetilde{\Delta}^{2}+\Omega^{2}_{N}$
with $\Omega_{N}=\sqrt{4N\mu\mu^{*}}=\Omega_1\sqrt{N}$ being the
"stimulated" Rabi frequency. From the associated eigenstates, it
is easy to show that $\mid \psi_{N}(t)\rangle=e^{-iHt}\mid
\psi_{N}(0)\rangle$ reads $\sum_{p=(0,1)} a_{N}^{(p)}(t)\mid
N-p,X^p\rangle$, within an irrevelant phase factor
$e^{-i(N\omega_p+\Delta/2)t}$. The prefactors $\alpha_N^{(p)}$(t)
are given by
\begin{equation}\label{eq3}
a_{N}^{(0)}(t)=e^{-\gamma t/2}
\left[cos(\widetilde{\Omega}_{N}t/2)+i(\widetilde{\Delta}/\widetilde{\Omega}_{N})
sin(\widetilde{\Omega}_{N}t/2)\right]
\end{equation}
\begin{equation}\label{eq4}
a_{N}^{(1)}(t)=-i e^{-\gamma t/2}
(2\mu^{*}\sqrt{N}/\widetilde{\Omega}_{N})
sin(\widetilde{\Omega}_{N}t/2)
\end{equation}
in agreement with textbooks \cite{ref9}.

The photon number, equal to $\langle\psi_{N}(t)\mid A^{+}A\mid
\psi_{N}(t)\rangle$, thus reads for a two-level atom initially in
its ground state
\begin{equation}\label{eq5}
\mathcal{N}_a(t)=N[P_{N}^{(0)}(t)+P_{N}^{(1)}(t)]-P_{N}^{(1)}(t)
\end{equation}
where $P_{N}^{(p)}(t)=\mid a_{N}^{(p)}(t)\mid^{2}$ are the
probabilities for the atom to have no excitation (p=0) or one
excitation (p=1). Using eqs.(\ref{eq3},\ref{eq4}), these
probabilities are
\begin{equation}\label{eq6}
P_N^{(1)}(t)=\frac{e^{-\gamma
t}}{2}\frac{\Omega_N^2}{\mid\widetilde{\Omega}_{N}^2\mid}[ch(t
O'_N)-cos(t O_N)]
\end{equation}
\begin{equation}
P_{N}^{(0)}(t)=e^{-\gamma t}\left[I_{N}^{+}ch(t O'_N)-D_N
sh(tO'_N) +I_N^{-}cos(t O_N)-D^{'}_N sin(t O_N) \right]\label{eq7}
\end{equation}
where we have set $\widetilde{\Omega}_{N}=O_N+iO'_N$,
$\widetilde{\Delta}/\widetilde{\Omega}_N=D_N+iD^{'}_N$ and
$I_N^{(\pm)}=(1\pm
|\widetilde{\Delta}^2/\widetilde{\Omega}_N^2|)/2$.
\begin{figure}
\begin{center}
\mbox{\includegraphics[height=5.5 cm,width=9 cm]{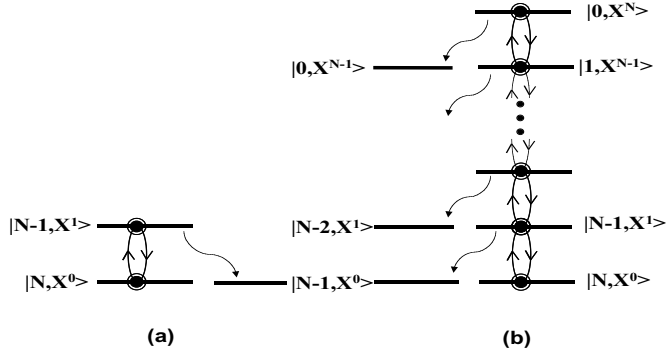}}
\caption{Possible couplings of a two-level atom, (a), or a
semiconductor, (b), induced by the presence of N photons, when the
excited states have external relaxation (wavy lines).)
}\label{fig_abs}
\end{center}
\end{figure}

\section{Semiconductor polariton}
The problem of photon interaction with a semiconductor has
similarities with the one of a two-level atom except that, after
the transformation of one photon into one exciton, it is a priori
possible to excite another electron from the valence band, and,
again, another one, until all the photons are transformed into
excitations. This of course implies that the photon number N is
small compared to the total number of valence electrons. In order
to possibly describe this phenomena in a simple way, this also
implies that N is not too large to end with a set of N excitons,
which can be considered as non-interacting, i.e., all at the same
energy. This is valid if the Coulomb and Pauli interactions
\cite{ref12,ref13} between N excitons are small, i.e., if
$N(a_x/L)^d\ll 1$, $a_x$ and $L$ being the exciton and sample
size, while $d$ is the space dimension. In this limit, the
excitons can be taken as non-interacting bosons \cite{noteb}. The
coupled semiconductor-photon hamiltonian then reads as
eq.(\ref{eq1}), where $B^+$ is now the ground state exciton
\cite{ref14} creation operator. It is such that $B^+\mid
X^{n}\rangle=\sqrt{n+1}\mid X^{n+1}\rangle$ for any $n\geq 0$,
with $\mid X^{n}\rangle=(n!)^{-1/2}(B^+)^n\mid v\rangle$ being the
normalized semiconductor state with $n$ boson-excitons.

If we consider an initial state similar to the one of a two-level
atom, namely \newline $\mid \psi_{N}(t=0)\rangle=\mid N,X^{0}
\rangle$, with N photons and the semiconductor in its ground
state, the hamiltonian (\ref{eq1}) couples it to $\mid N-1,X^{1}
\rangle$,...,$\mid 0,X^{N}\rangle$, with possible relaxation to
states like $\mid N-1,X^{0}\rangle$ if the exciton has external
couplings, i.e., if $\gamma$$\neq$0 (see fig. (1b)).

In order to obtain the time evolution of $\mid
\psi_{N}(t)\rangle$, we can either diagonalize $H$ in this (N+1)
subspace or better, rewrite $H$ in a diagonal form. As for
$\gamma\neq 0$, $H$ is not hermitian, this diagonalization is less
trivial than the one for textbook polaritons because, $H$ and
$H^+$ being different, they have different eigenstates. It is
however easy to check, just by replacing, that $H$ given in
eq.(\ref{eq1}), also reads
\begin{equation}\label{eq8}
H= E_1^{(-)}\alpha^+_k\alpha_b+E_1^{(+)}\beta_k^+\beta_b
\end{equation}
where $E_1^{(\pm)}$ are the eigenenergies for N=1 photon, given in
eq.(\ref{eq2}). The operators $\alpha^+_k$ and $\beta_k^+$ are
defined as
\begin{equation}
\alpha^+_k=\frac{(\widetilde{\Delta}+\widetilde{\Omega}_{1})A^+-2\mu^{*}B^+}{\sqrt{(\widetilde{\Delta}+\widetilde{\Omega}_{1})^2+\Omega_{1}^2}}\hspace{1cm}
\beta_k^+=\frac{(\widetilde{\Delta}+\widetilde{\Omega}_{1})B^++2\mu
A^+}{\sqrt{(\widetilde{\Delta}+\widetilde{\Omega}_{1})^2+\Omega_{1}^2}}\label{eq10}
\end{equation}
while $\alpha_b$ and $\beta_b$ read as $\alpha_k$ and $\beta_k$
with ($\widetilde{\Delta}^*+\widetilde{\Omega}_1^*$) replaced by
($\widetilde{\Delta}+\widetilde{\Omega}_1$). For a finite exciton
broadening, i.e., $\widetilde{\Delta}\neq\widetilde{\Delta}^*$,
the operators ($\alpha^{+}_k,\alpha_b$) and ($\beta_k^+,\beta_b$)
appearing in $H$ are not conjugate. They however are the ones
which fulfill the commutation relations
\begin{equation}\label{eq11}
\left[\alpha_b,\alpha^{+}_k\right]=1=\left[\beta_b,\beta_k^+\right]\hspace{1cm}
\left[\alpha_b,\beta_k^+\right]=0=\left[\beta_b,\alpha^{+}_k\right]
\end{equation}
so that $(H-E_1^{(-)})$$\alpha_k^+$$|v>$= 0 while
$<v|\alpha_b$$(H-E_1^{(-)})$= 0 and similarly for $\beta_k^+$ and
$\beta_b$. The operators $\alpha_k^+$ and $\beta_k^+$ have thus to
be seen as the creation operators for polaritons dressed by
exciton relaxation in the ket space, while $\alpha_b$ and
$\beta_b$ are the ones in the bra space.

To our best knowledge, the diagonal form of the coupled
photon-exciton hamiltonian written in eq.(\ref{eq8}), has not been
given before. The {\it generalized polariton operators} in the bra
and ket space it contains, are however of conceptual importance to
possibly describe photon-exciton interaction with relaxation.

By writing the photon creation operator $A^+$ in terms of
$\alpha^{+}_k$ and $\beta_k^+$, we can show, from the commutation
relations (\ref{eq11}) and the expression of the hamiltonian given
in eq.(\ref{eq8}), that the initial state $\mid
N,X^{0}\rangle=(N!)^{-1/2}(A^+)^N\mid v\rangle$, with $\mid
v\rangle$ being the exciton and photon vacuum, transforms into
$\mid\psi_N(t)\rangle=\mid\Phi_N(1,1;t)\rangle$ with
\begin{eqnarray}
\mid\Phi_N(x,y;t)\rangle=(N!)^{-1/2}\left(A^+_{x,y;t}\right)^N\mid
v\rangle
\label{eq12}\\
A^+_{x,y;t}=x a_1^{(0)}(t)A^++y a_1^{(1)}(t)B^+\label{eq13}
\end{eqnarray}
within an irrevelant phase factor $e^{-i(\omega_p+\Delta/2)t}$.
$a_1^{(0)}(t)$ and $a_1^{(1)}(t)$ are the prefactors appearing in
the time evolution of a two-level atom in the presence of N=1
photon, as given in eqs.(\ref{eq3},\ref{eq4}).

From this compact expression of $\mid\psi_N(t)\rangle$, it is easy
to obtain the time evolution of the photon number
$\langle\psi_N(t)\mid A^+A\mid\psi_N(t)\rangle$. Indeed, as
$A^+A\mid\psi_N(t)\rangle$ is just
$\partial_{x}\mid\Phi_{N}(x,y;t)\rangle$ taken for $x=y=1$, this
photon number is nothing but
(1/2)$\partial_{x}\langle\Phi_{N}(x,y;t)\mid\Phi_{N}(x,y;t)\rangle\mid_{x=y=1}$.
In order to get the norm of $|\Phi_{N}(x,y;t)\rangle$, we can note
that
\begin{equation}\label{eq14}
A_{x,y;t}\left(A^+_{x,y;t}\right)^N\mid v\rangle=N b
_{x,y;t}\left(A^+_{x,y;t}\right)^{N-1}\mid v\rangle
\end{equation}
where we have set
\begin{equation}\label{eq15}
b_{x,y;t}=[A_{x,y;t},A^+_{x,y;t}]=x^2P_1^{(0)}(t)+y^2P_1^{(1)}(t)
\end{equation}
$P_1^{(0,1)}(t)$ being the probabilities given in
eqs.(\ref{eq6},\ref{eq7}), for N=1 photon. This shows that the
norm of $|\Phi_{N}(x,y;t)\rangle$ is nothing but $b^N_{x,y;t}$, so
that the photon number for a semiconductor initially in its ground
state, finally reads
\begin{equation}
\mathcal{N}_{sc}(t)=(N/2) \hspace{.2cm}
(b_{x,y;t})^{N-1}\partial_{x}b_{x,y;t}\mid_{x=y=1}
=NP_1^{(0)}(t)\left(P_1^{(1)}(t)+P_1^{(0)}(t)\right)^{N-1}\label{eq16}
\end{equation}
This nicely compact analytical result is valid for any time, any
photon number, any detuning, any exciton broadening, and any
photon-semiconductor coupling.

\section{Time evolution of the photon number}

The time evolution of the photon number for an atom or a
semiconductor initially in their ground state are given by
eqs.(\ref{eq5},\ref{eq16}). These analytical expressions can be
used to obtain the photon number for any given set of experimental
conditions. We are just going here to discuss the limiting cases
of importance for physical understanding.

{\it (i) In the absence of matter-photon coupling}, i.e., for
$\Omega_1=0$, we have $P_N^{(1)}(t)=0$ while $P_N^{(0)}(t)$=1: the
matter stays in its ground state, and the photon number stays
equal to N, as expected.

{\it (ii) In the absence of excited state relaxation}, i.e., for
$\gamma=0$, the sum $P_N^{(1)}(t)+P_N^{(0)}(t)$ stays equal to 1.
The photon number change n(t)=N-$\mathcal{N}$(t) reduces to
$P_N^{(1)}(t)=(\Omega_{N}^2/\omega_N^2)sin^2(\omega_N t/2)$ in the
case of atom and to
N$P_1^{(1)}(t)=N(\Omega_{1}^2/\omega_1^2)sin^2(\omega_1t/2)$ in
the case of semiconductor, with $\omega_N^2$ being
$\Omega_N^2+\Delta^2$. At resonance, i.e., $\Delta$=0, we recover
that for a dressed atom, the photon number oscillates, between N
and (N-1), at the {\it stimulated Rabi frequency} $\Omega_N$. For
a semiconductor, it oscillates, at the vacuum Rabi frequency
$\Omega_1$, between N and 0, so that all the photons can possibly
be transformed into excitons. When the detuning increases, the
frequency of these oscillations increases, while the photon number
stays closer to N.

{\it (iii) When the excited state relaxation is included}, i.e.,
for $\gamma\neq 0$, the sum $P_N^{(1)}(t)+P_N^{(0)}(t)$ decreases
with time. For $\gamma$ small, the Rabi oscillations are
essentially damped, while with increasing $\gamma$, the decrease
of $P_N^{(1)}(t)+P_N^{(0)}(t)$ ends by controlling  the photon
number change. For $\gamma\gg\Omega_N$, the expansion of
eqs.(\ref{eq6},\ref{eq7}) in
$\eta_N^2=\Omega_N^2/(\gamma^2+\Delta^2)$, shows that, as
$\widetilde{\Omega}_N\simeq
\widetilde{\Delta}+\widetilde{\Delta}^*\eta^2_N/2$
\begin{equation}
P_N^{(1)}(t)\simeq (\eta_N^2/4)e^{-\gamma\eta_N^2t/2}\hspace{.5cm}
P_N^{(0)}(t)\simeq
\left(1+\frac{\eta_N^2}{2}\frac{\gamma^2-\Delta^2}{\gamma^2+\Delta^2}
\right)e^{-\gamma\eta_N^2t/2}\label{eq17}
\end{equation}
for $\gamma t \gg 1$. Using the above equations, we find that, for
$\gamma \gg (t^{-1}, \Omega_N)$, the photon number of a two-level
atom, given in eq.(\ref{eq5}), tends to $\mathcal{N}_a(t)\simeq
Ne^{-\gamma\eta_N^2t/2}$ while the one for a semiconductor, given
in eq.(\ref{eq16}), tends to $\mathcal{N}_{sc}(t)\simeq
Ne^{-N\gamma\eta_1^2t/2}$. Since $\eta_N^2=N\eta_1^2$, this shows
that, in this limit,
$\mathcal{N}_a(t)\simeq\mathcal{N}_{sc}(t)\simeq Ne^{-t/T}$ with
$T$ given by
\begin{equation}\label{eq18}
\frac{1}{T}=\frac{N\gamma\eta_1^2}{2}=2\pi\mid\mu\sqrt{N}\mid^2
\widehat{\delta}_\gamma(\Delta)
\end{equation}
$\widehat{\delta}_\gamma(\Delta)=\gamma/\pi(\Delta^2+\gamma^2)$
being a delta "function" of width $\gamma$.

We see that, for $\gamma$ large compared to $t^{-1}$ and
$\Omega_N$, the photon number for a two-level atom and a
semiconductor, both behave as $Ne^{-t/T}$, the characteristic time
of this exponential decay following a kind of Fermi Golden rule:
$\mu^*\sqrt{N}$ is indeed the matrix element of the matter-photon
coupling between the initial state, N photons and the matter in
its ground state, and the excited state, (N-1) photon and one
excitation. The energy conservation it contains is however
conceptually different from the standard one: It is linked to the
energy uncertainty induced by the excited state broadening, not
the one coming from the Heisenberg principle, as in the usual
Fermi Golden rule. The ground state of an atom or a semiconductor
being not coupled to a continuum but to a discrete state, the
exponential decay of the photon number does not come from
destructive interferences between Rabi oscillations, but from the
broadening of the excited state itself, in a non trivial way: It
is not a bare $e^{-\gamma t}$, but results from the interplay
between $\gamma$ and $\Omega_N$. This decay only exists when the
excited state energy uncertainty $\gamma$ is much larger than the
strength of the stimulated matter-photon coupling. If we see the
broadened excited state as a continuous set of states, all at the
same energy, we can say, in a very crude way, that the delta
"function" $\widehat{\delta}_\gamma$, which appears as a
multicative factor in the decay rate, "selects" a part of this
broadened excited state "continuum", in the same way as the delta
"function", $\delta_t$, of the usual Fermi Golden rule, "selects"
the part of the continuum coupled to the ground state.

{\it Conclusion}

We have derived the time evolution of the photon number in the
presence of an atom or semiconductor in their ground state, using
the same framework. This led us to introduce generalized polariton
operators dressed by the exciton finite lifetime, which differ in
the bra and ket spaces. They actually are of conceptual importance
for the study of photon-semiconductor interaction in the presence
of relaxation.

When the photons couple the ground state to a {\it discrete
excited state which has an infinite lifetime}, we recover that not
only a single two-level atom, but also a {\it macroscopic}
\cite{notec} semiconductor have Rabi oscillations,  at the
stimulated frequency, $\Omega_1\sqrt{N}$, for atoms, and at the
vacuum one, $\Omega_1$, for  semiconductors.

On the opposite, when the {\it excited state lifetime is small},
the photon number has the {\it same} exponential decay for atom
and semiconductor, with a characteristic time given by the Fermi
Golden rule. This rule can indeed be extended to {\it transitions
to a discrete state}, provided that the discrete state is highly
broadened compared to the stimulated matter-photon coupling, and
that the "energy conservation" this Fermi Golden rule contains, is
enforced at the scale of the excited state broadening.


\begin{thebibliography}{99}
\bibitem{b} Coupling with phonons scatters the photocreated excitons to other exciton states;
The semiconductor stays excited while, for a two-level atom, the
electron returns to its ground state by fluorescent emission.
\bibitem{notea} With increasing excitation, these
excitons may end by going out the strong coupling limit, due to
interaction between excitons which decreases their lifetime.
\bibitem{noteb} The theory of interacting polaritons using the ideas
of references \cite{ref12} and \cite{ref13} is far beyond the
scope of this paper, due to the quite tricky close-to-boson
character of the exciton.
\bibitem{notec} Rabi oscillations seen in quantum dots \cite{ref10} are
actually similar to the ones of a two-level atom because in a dot,
a given excited state can be occupied by one electron only (if we
forget spin) while in an extended semiconductor, many electrons,
all at the same energy, can be created.
\end{thebibliography}
\end{document}